\documentstyle[twoside,fleqn,espcrc2]{article}

\newcommand{\AMS}{{\protect\the\textfont2
 A\kern-.1667em\lower.5ex\hbox{M}\kern-.125ems}}

\title{{\Large Why {\em two} makes it more 
attractive than just with one...} \\  or  \\
Bianchi class-A models and Reissner-Nordstr\"om 
Black Holes in Quantum N=2 Supergravity\thanks{Extended 
version of talks/communications  presented at the 
conferences 
{\em Constrained Dynamics and Quantum Gravity}, 
Sta. Marguerita, September 1996, Italy;   
{\em New Voices in Gravitation and Quantum Gravity}, Penn State, 
November 1996, USA and
{\em 18th Texas Symposium on Relativistic Astrophysics}, December 
1996, Chicago, USA}}
\author{{\sf P.V. Moniz} \address{Department of 
Applied Mathematics and Theoretical Physics \\ 
University of Cambridge\\  Silver Street, Cambridge,  CB3 9EW \\
 United Kingdom}%
\thanks{This work was supported by the  JNICT/PRAXIS XXI Fellowship
 BPD/6095/95 (Sub-Programa Ci\^encia e Tecnologia do $2^{\rm o}$ 
Quadro Comunit\'ario de Apoio) and the project PBIC/P/MAT/2150/954; 
e-mail: {\sf prl\-vm\-10\-@am\-tp.c\-am.ac.uk; P\-MO\-NIZ@Delphi.com}; 
URL: {\sf http://www\-.dam\-tp.\-c\-am.\-ac.uk/us\-er/\-pr\-lv\-m\-10}}}

\begin{document}

\begin{abstract}
Bianchi class-A models and Reissner-Nordst\"rom (RN) black hole scenarios 
are considered from the point of view of quantum {\em N=2} Supergravity.
It is shown that 
the presence of  Maxwell fields in the supersymmetry 
constraints implies a non-conservation of the fermion number 
present in Bianchi class-A models. This effect 
corresponds to a mixing between different 
(Lorentz invariant) fermionic sectors in the wave function of the Universe. 
Quantum states 
are constituted by  exponentials of N=2 
supersymmetric 
Chern-Simons functionals. 
With respect to the RN case, we 
analyse some problems and features present in a 
reduced  model with 
supersymmetry. Lines of subsequent research work 
are then provided.
\end{abstract}

\maketitle

\section{Introduction}

\indent

$N=2$ supergravity \cite{PvN,3,7} realises  Einstein's dream of 
unifying gravity with electromagnetism. The theory contains 
2 real (1 complex) gravitino besides the gravitational (tetrad) and Maxwell fields. 
It was in this theory that finite probabilities for loop diagrams with gravitons were first obtained. In particular, the photon-photon scattering process which is divergent in 
a Einstein-Maxwell theory, was shown to be finite when $N=2$ supergravity was 
considered (cf. \cite{PvN} and references therein, \cite{8a}).

Another  important property 
is the fact that N=2 supergravity has {\em more} symmetries  \cite{PvN}
 which rotate fermionic fields into themselves 
via $O(2)$ transformations. 
Gauging the $O(2)$ symmetry 
 we get a set of coupling constants. 
The presence of  these local symmetries proved to be crucial in accomplishing 
 the 
finite results present in \cite{8a}. 
Hence, the 
additional symmetries and  elegant coupling between  several physical 
variables existing  in $N=2$ supergravity 
convey indeed  to an attractive scenario to investigate.

In this work we will consider the theory of  $d=4, N=2$  supergravity from an 
ADM  metric representation point of view. We will analyse 
 Bianchi class-A models \cite{a,b,c} and Reissner-Nordstr\"om (RN) 
black holes \cite{U}. 
In section 2 we address the canonical 
formulation for the case of Bianchi class A models \cite{a,b}, 
while an outlook on the RN case is  included in section 3. 
This paper is then concluded in section 4 
 with our conclusions and a discussion on  subsequent lines of research.

\section{ Bianchi Models in $N=2$ Supergravity}

\vspace{0.2cm}

{\tt [This section is based on joint work with A.D.Y.Cheng \cite{a,b}]}

\vspace{0.2cm}

\indent 

The action for the N=2 supergravity  theory (with a 
$O(2)$ global symmetry) can be written as \cite{PvN}
\begin{eqnarray}
{\cal L} = & - & \frac{e}{2 \kappa^2} R[e^a_\mu,\omega(e^a_\mu, \psi^{(a)}_\nu)] 
\nonumber \\ &-&  
{e \over 2} \bar 
\psi^{(a)}_{\mu} \epsilon^{\mu \nu\rho \sigma} \gamma_5 \gamma_\nu
D_{\rho}(\omega) \psi^{(a)}_{\sigma} \nonumber \\
& - &  {e \over 4} F^2_{\mu \gamma} + {\kappa \over 4 \sqrt 2} 
\bar \psi^{(a)}_{\mu}\left[e (F^{\mu \nu} + \hat F^{\mu \nu}) 
\right. \nonumber \\ &
+& \left.  {1 \over2} \gamma_5({\tilde F^{\mu \nu}}
+ \tilde {\hat F}{}^{\mu \nu})\right] \psi^{(b)}_{\nu} \epsilon^{ab},
\end{eqnarray}
where 
\begin{equation}
\hat F_{\mu \nu} = \left( \partial_{\mu} A_{\nu} - {\kappa \over 2 \sqrt 2} \bar \psi^{(a)}_{\mu}
\psi^{(b)}_{\nu} \epsilon^{ab} \right) - (\mu \leftrightarrow \nu).
\end{equation}
and $\tilde {F}_{\mu \nu}$ equals $\epsilon_{\mu \nu \rho \sigma} F^{\rho \sigma}$.
In the metric representation, the independent variables are taken to be the 
tetrad components $e^a_\mu$, where $a=1,..,4$ are Lorentz indices and $\mu = 0,..,3$ are Einstein indices, the gravitinos $\psi^{(a)}_\mu$ and their Hermitian conjugate (here depicted in 4-component representation;  $(a)=1,2$ are $O(2)$ group indices) and $A_\mu$ is a Maxwell field. $\kappa$ is the gravitational 
constant, $\omega$ is the connection, 
$\gamma_\mu$ are Dirac matrices and $\gamma_5 = \gamma_0 \gamma_1 \gamma_2 \gamma_3$. 
Furthermore, $\epsilon^{12} = 1, \epsilon^{21} = -1$.

{\it En route} to the  canonical quantization of Bianchi class A models,  we require 
the following two steps to be complied \cite{a,b,c}. 
 On the one hand, we ought to re-write the action (1) in 
2-component spinor notation. We do so using the conventions in \cite{c}. 
On the other hand, we impose a consistent Bianchi Anzatse for all fields.

The supersymmetry constraints take the  form
\begin{eqnarray}
\bar S^{(a)}_{A'} & = &  -i p^{i}_{AA'} \psi^{(a)A}_{i} + 
\epsilon^{ijk} e_{AA'i} ~\omega^{(3s)}_j \psi^{A(a)}_k
\nonumber \\
 & + & \kappa \epsilon^{ab} \left[ \pi^i - i h^{\frac{1}{2} } {\kappa \over 4} 
\epsilon^{imn} \epsilon^{cd} \psi_m^{(c)B} \psi_{nB}^{(d)} 
\right. \nonumber \\ &
+& \left.  i h^{\frac{1}{2} } \frac{1}{8} F_{jk} \epsilon^{ijk}  \right] \bar \psi^{(b)}_{iA'},
\label{eq:susyconbar}\end{eqnarray}
where $p^i_{AA'}$ is the momentum conjugate to the tetrad, 
$
\pi^i$ 
is the momentum associated with the Maxweel field. 
Notice that the contribution 
from the Maxwell field changes according to the chosen Bianchi type due to 
the dependence of $F_{jk}$ on the structure constants of the Bianchi group.



Choosing $\left(A_i, e_{iAA'}, \psi^{(a)A}_i\right)$ as the 
 coordinates variables in our 
minisuperspace, 
our canonical relations will be
\begin{eqnarray}
\left[ e_i^{~AA'}, {\hat p}^j_{BB'} \right]  & = &
 i \hbar \delta_i^j \delta_B^A \delta_{B'}^{A'},~~\\
\left[ \psi_{i}^{(a)A}, \bar \psi_{j}^{(b)A'} \right] 
 & = &  -i \hbar \delta^{ab} D^{AA'}_{~~ij} ,~~\\
\left[ A_i, \pi^j \right]  & = &   i \hbar \delta_i^j,
\end{eqnarray}
which imply
\begin{eqnarray}
\bar \psi_l^{(a)A'} &\rightarrow &  -i  D^{AA'}_{~~ij} \left(e^{AA'}_\mu, 
 \psi^{A(b)}_\mu \right) {\partial \over \partial \psi_i^{(a)A}},\\ ~
{\hat p}^j_{AA'}  &\rightarrow &  -i  {\partial \over \partial e_j^{AA'}};~
\\   \pi^i  &\rightarrow & -i  {\partial \over \partial A_i}
\end{eqnarray}
and consequently, 
we obtain the quantum supersymmetry constraints
\begin{eqnarray}
 \bar S^{(a)}_{A'} & = & \psi^{(a)A}_i {\partial \over \partial e_i^{AA'}} 
+ \epsilon^{ijk} e_{AA'i} \omega^{(3s)}_j \psi^{A(a)}_k 
\nonumber \\ &+& 
 {1\over 2} \epsilon^{ijk} \Gamma^{ab} \psi_i^{(a)A} 
\psi^{(b)}_{jA} D^B_{A'lk} 
{\partial \over \partial \psi_l^{(b)B}} \nonumber\\
 & + &  \epsilon^{ab} D^C_{A'mi} \left[  {\partial \over \partial A_i} 
- {e \over 4}
\epsilon^{ijk} \epsilon^{cd} \psi_j^{(c)B} \psi_{kB}^{(d)} \right.
\nonumber \\ 
&+& \left.  \frac{i}{8}e \epsilon^{ijk} F_{jk}
\right]
 {\partial \over \partial \psi_m^{(b)C}},
\end{eqnarray}
with  $S^{(a)}_A$ is the Hermitian conjugate, 
 $\Gamma^{12} = \Gamma^{21} = 1$, the remaining are zero and 
\begin{equation}
D^{AA'}_{~~ij} = -2i h^{-{1 \over2 }} e_{~~j}^{AB'} e_{iBB'} n^{BA'}.
\end{equation}
The factor ordering of the supersymmetry constraints has 
been chosen such that they 
give the correct left and 
right superysmmetry transformations.

We now address the physical states which are solutions of 
the above constraints. The quantum states may be described by the 
wave function $\Psi (e_{AA'i}, A_j, \psi^{(a)}_{Ai})$. Lorentz invariance 
implies  that the wave function 
ought to be  
 expanded in even powers of $\psi^{(a)}_{Ai}$, symbolically 
represented by $\psi^0, \psi^2$ up to $\psi^{12}$.

An important consequence 
is that {\it neither} of  the supersymmetry constraints 
$\bar S_{A'}^{(a)}$ and $S_A^{(a)}$ conserves 
fermion number and hence causes a mixing between 
fermionic modes in $\Psi$. This is due to  terms involving  ${\partial \over \partial A_i} 
{\partial \over \partial \psi_m^{(b)C}}$
and ${\partial \over \partial A_i} \psi^{(b)}_{iA}$ or the ones associated with $F_{jk}$. In fact, while the remaining fermionic 
terms in  $\bar S_{A'}^{(a)}$ by acting on $\Psi$ 
effectively increase the fermionic order by a factor of one, ${\partial \over \partial A_i} 
{\partial \over \partial \psi_m^{(b)C}}$ decrease 
it by the same amount. Concerning the $S_A^{(a)}$ 
constraint,  the situation is precisely the reverse.

The nature of this problem can also be better understood 
 as follows. Let us consider the two fermion level whose 
 more general Ansatz for the  wave function is 
\begin{eqnarray}
\Psi_2 &=& 
 \left( C_{ijab} + E_{ijab} \right) \psi^{(a)iB} \psi^{(b)j}_{~~~B}
\nonumber \\ &+& 
 \left( U_{ijkab} + V_{ijkab} \right) e^i_{AA'} n_B^{A'} \psi^{(a)jA} \psi^{(b)kB}
\end{eqnarray}
where $C_{ijab} = C_{(ij)(ab)}$, $E_{ijab} = E_{[ij][ab]}$, $U_{ijkab} = U_{i(jk)[ab]}$ and
$V_{ijkab} = V_{i[jk](ab)}$.
Take $ \bar S^{(a)}_{A'} \Psi = 0$, 
$S^{(a)}_{A} \Psi = 0$, where $\Psi$ is truncated 
in the second order of the fermionic Lorentz invariant sector, i.e., $\Psi  = \Psi_0 + 
\Psi_2$, with 
$\Psi_2$ given as above. We obtain a set of equations where the ones linear in the 
fermionic variables relate terms such as $\frac{\partial \Psi_0}{\partial a_i}$ with 
 $\frac{\partial C_{ijab}}{\partial A_i}$, $\frac{\partial E_{ijab}}{\partial A_i}$, 
$\frac{\partial U_{ijkab}}{\partial A_i}$, $\frac{\partial V_{ijkab}}{\partial A_i}$
(from $\bar S^{(a)}_{A'}$) 
and 
 $\frac{\partial C_{ijab}}{\partial a_i}$, $\frac{\partial E_{ijab}}{\partial a_i}$, 
$\frac{\partial U_{ijkab}}{\partial a_i}$, $\frac{\partial V_{ijkab}}{\partial a_i}$
 with 
$\frac{\partial \Psi_0}{\partial A_i}$ 
(from $ S^{(a)}_{A}$). Here $a_i, i=1,2,3$,
stand for scale factors in a Bianchi class-A model, $\Psi_0$ is the 
bottom (bosonic) sector (in the expansion of $\Psi$ into Lorentz invariant fermionic 
sectors) and $\Psi_0, C_{ijab}, E_{ijab}, U_{ijkab}, V_{ijkab}$ 
are functions of $A_j, a_i$ solely.

As we can conclude, the spin-1 field 
terms in the supersymmetry constraints impose a mixing (or coupling) between 
different fermionic sectors in $\Psi$. However, the present situation is rather different 
form the one in Friedmann-Robertson-Walker (FRW) models in $N=1$ supergravity, 
when only  scalar fields and fermionic partners are present (cf. ref. \cite{c}). There is a 
{\it particular} mixing in this case as well but only within each fermionic level (say, 
second order in fermionic terms in $\Psi$) and decoupled from other Lorentz invariant 
fermionic  sectors with different order. In the present case, the mixing of fermionic sectors is between  fermionic sectors of the same and {\it any}  different adjacent order. 

It is tempting to see the relation between 
(simbolical) gradient  terms such as $\frac{\partial \Psi_0}{\partial a_i}$ with 
 $\frac{\partial \Psi_2}{\partial A_i}$ (from $\bar S^{(a)}_{A'}$) and 
 $\frac{\partial \Psi_2}{\partial a_i}$ with 
$\frac{\partial \Psi_0}{\partial A_i}$ (from $ S^{(a)}_{A}$), like others pointed out 
above, 
as a consequence of $N=2$ supergravity \cite{PvN},  which realizes Einstein's 
dream of 
unifying gravity with electromagnetism. These relations establish a kind of {\em 
duality} between the bosonic coefficients in fermionic sectors of adjacent order 
relatively  to derivatives  with respect to gravitational and Maxwell degrees of freedom 
$\left(\frac{\partial}{\partial A_i}, \frac{\partial}{\partial a_i}\right)$. These 
derivatives are intertwined on an 
equal ground as far as $\Psi$ (and its bosonic coefficients) are concerned.

One physical consequence of considering  $O(2)$ local invariance, in spite of the 
additional difficulties caused now by 
 cosmological constant and gravitinos mass-terms \cite{PvN,3,7}, 
 is to allow us 
to extract some information concerning the form of the wave function. 
From a somewhat simplified adaptation of the method  outlined in ref. \cite{F}. 
we will 
get \cite{a,b,c}:
\begin{equation} 
\Psi_{12} \sim \Sigma_{m=1,2,3} (A^m)^2 e^{\epsilon^{ijk}A_i F_{jk}},
\end{equation}
where $A^2 = A_i A^i$ and 
\begin{equation}
A^4 e^{-(a_1^2 + a_2^2 + a_3^2) + (a_1 a_2 + a_2 a_3 + a_3 a_1)} e^{\epsilon^{ijk}A_i F_{jk}}, 
\end{equation}
\begin{equation}
A^4 e^{-(a_1^2 + a_2^2 + a_3^2) + (a_1 a_2 - a_2 a_3 - a_3 a_1)} e^{\epsilon^{ijk}A_i F_{jk}},
\end{equation}
could exist in the 8-fermion sector, half for each gravitino type
(see also ref. \cite{a,b,c,12} and references therein).

\section{Reissner-Nordstr\"om Black Holes}

\vspace{0.2cm}

{\tt [Research work outlined in this section is currently in progress 
and some in colaboration with Jarmo M\"akel\"a]}

\vspace{0.2cm}

The canonical 
quantization of Reissner-Nordstr\"om black holes in N=2 supergravity was 
motivated by previous research on: $(i)$ 
canonical quantization of black holes in general relativity \cite{T}--\cite{T4};
$(ii)$ quantization of minisuperspaces in N=1 and N=2 supergravity \cite{c};
$(iii)$ Analisys of (classical) black hole solutions in N=2 
and N=4 supergravity \cite{9}.

A possible approach  would be to take for the tetrad \cite{T3,T4} 

\[ e^a_\mu = \left( \begin{array}{cccc}
 -N(t,r) & 0 & 0 & 0 \\
\Lambda N^r (t,r) & \Lambda (t,r) & 0 & 0 \\
0 & 0 & R(t,r) & 0 \\
0 & 0 & 0 & R \sin \theta
\end{array}
\right),
\]
while for the Maxwell and gravitino fields we could 
use 
\begin{eqnarray} 
 A_\mu & \mapsto & 
\left( 
\Phi (t,r), \Gamma (t,r), \right. \nonumber \\
 \hat A_\theta & \sim & A_\theta (t,r) \hat f(\theta,\phi), 
\nonumber \\ 
\tilde A_\phi & \sim & \left. A_\phi (t,r)  \tilde f(\theta,\phi) \right),
\end{eqnarray}
\begin{eqnarray}
\psi^{(a)A}_\mu & \mapsto &
\left( 
\psi^{A(a)}_0 (t,r), 
\psi^{A(a)}_r (t,r), \right.
\nonumber \\ 
\hat \psi^{A(a)}_\theta  & \sim & \psi^{A(a)}_\theta 
(t,r) \hat g(\theta,\phi), 
\nonumber \\  
\tilde \psi^{A(a)}_\phi & \sim & \left.
\psi^{A(a)}_\phi (t,r)  \tilde g (\theta,\phi) \right),
\end{eqnarray}
where the functions $\tilde f (\theta,\phi)$,
$\tilde g (\theta,\phi), 
$ $\hat f (\theta,\phi)$, $ 
\hat g (\theta,\phi)$ 
are still to be identified.

The supersymmetry quantum constraints
will then have the  {\em formal} expression 
\begin{eqnarray}
\bar S^{(a)}_{A'} & = &  -i p^{i}_{AA'} \psi^{(a)A}_{i} + 
\epsilon^{ijk} e_{AA'i} [\Lambda, R]  \omega^{(3s)}_j \psi^{A(a)}_k
\nonumber \\
 & + & \kappa \epsilon^{ab} \left[ \pi^i - i h^{\frac{1}{2} } {\kappa \over 4} 
\epsilon^{imn} \epsilon^{cd} \psi_m^{(c)B} \psi_{nB}^{(d)} 
\right. \nonumber \\ &+& \left.  i h^{\frac{1}{2} } \frac{1}{8} F_{jk} 
[A_\theta, A_\phi] 
 \epsilon^{ijk}  \right] \bar \psi^{(b)}_{iA'}.
\nonumber \\
& + & 
\frac{1}{2} \epsilon^{ijk} e_{AA'i}[\Lambda, R] 
\partial_j \psi^{A(a)}_k
\nonumber \\
& + & \frac{1}{2} \epsilon^{ijk} e_{AA'i} [\Lambda, R] 
\omega^{0A}_{~Bj} [\partial_{\hat i} P(\Lambda, R)] \psi^{B(a)}_k 
\nonumber \\ &+&  
\frac{1}{2} \epsilon^{ijk} 
\psi^{A(a)}_i 
e_{AA'i} [\Lambda, R] 
\bar \omega^{kA'}_{~B'},
\end{eqnarray}
together with its Hermitian conjugate
while the gauge (Gauss) constraint is 
\begin{eqnarray}
Q & = & - ``N^{-1} \Lambda^{-1} R^2 (\partial_0 \Gamma - 
\partial_r \Phi)'' \nonumber \\
& + & 
\partial_i\left[ i\frac{NR^2 \Lambda \sin \theta}{2\sqrt{2}} 
\epsilon^{ijk} (\bar \psi^{(a)}_{jA'} \psi^{(b)A'}_{k} \right. \nonumber \\
& - & \left. 
\psi^{(a)A}_{j} \psi^{(b)}_{kA}) \epsilon^{ab}\right]
 \nonumber \\ 
& + & 
 \partial_i\left[ i\frac{NR^2 \Lambda \sin \theta}{\sqrt{2}} 
 (\bar \psi^{(a)0}_{A'} \psi^{(b)iA'} \right. \nonumber \\ 
&  +  & \left. 
\psi^{(a)0A} \psi^{(b)i}_{A}) \epsilon^{ab}\right].
\end{eqnarray}
From the term 
$\frac{1}{2} \epsilon^{ijk} e_{AA'i}[\Lambda, R] 
\partial_j \psi^{A(a)}_k $ in the supersymmetry 
constraint it seems that  that solutions could only be possible on the fermion 
sector with infinite number modes (see ref. \cite{DF} and also \cite{c}).

However, the use of a spherically symmetric spacetime metric in the study of 
Reissner-Nordstr\"om black holes in N=2 supergravity might be too 
restrictive. In fact, the black-hole metric obtained in \cite{aich}
is {\em not} spheriacally symmetric. This (classical) black hole 
solution in N=2 supergravity is a Grassmann-algebra-valued field which 
can be decomposed into a ``body'' (or component along unity) 
which takes value in the range of real or complex numbers and a 
``soul'' which is nilpotent. The ``body'' of the solution in \cite{aich} 
is spherically symmetric but the the ``soul'' (and hence the all metric solution)   
is not. Maybe one should include then the $N_\theta$ and $N_\phi$ 
components of the shift vector.

Nevertheless, a far more attractive 
scenario would be to follow K. Kuchar approach \cite{T3,T4} within the 
context present in \cite{aich}. 
In fact, we could aim in 
solving the constraints and, upon adequate boundary 
conditions, retrieve a {\em finite}-dimensional 
supersymmetric mechanical system and subsequently 
proceed with its quantization.

\section{Conclusions and {\em Outlook}}

\indent

Summarizing, 
we applied the theory of  $N=2$ supergravity 
to quantum  Bianchi class-A 
models and outlined its practice towards  quantum 
 Reissner-Nordstr\"om black holes. We
 found out that the presence of  Maxwell field 
terms in the supersymmetry constraints, 
will produce  a mixing between different Lorentz 
invariant fermionic sectors in the wave function of the universe.

 Semi-classically, the constraints can be solved for the 
Bianchi class-A case and the WKB 
wave function has the form of the exponential of the $N=2$ supersymmetric 
extended Chern-Simons functional. 

The main motivation to 
study  black holes  from the point of view of canonical 
quantization 
 is to go beyond a semi-classical description based in 
quantum fields in curved space-time. 
The Reissner-Nordstrom black-hole \cite{U} is  an important case to address. It 
corresponds to a charged black-hole and for a specific combination of its mass and 
charge, it is shown that its temperature is zero. These {\it extreme} cases  hold 
important supersymmetry properties, 
which are related to other issues like naked 
singularities and 
(in $N=2$ supergravity)
a Bogolmony bound for the black hole mass \cite{9}. Our objectives 
are  to study and analyse these features directly from a canonical quantization point of 
view and possibly infer more about the supersymmetry and thermodynamical 
properties of the (extreme) Reissner-Nordstrom black-holes, their
 entropy and the problem of time.


\vspace{0.05cm}

{\Large\bf Acknowledgements}

\vspace{0.05cm}

\indent 

The author 
is most grateful to P. Aichelburg, M. Cavaglia, A.D.Y. Cheng, P. Hajicek, 
S.W. Hawking, R. Kallosh, J. Louko and in particular 
J. M\"akel\"a for useful comments and discussions which led 
to additional motivation  on  the research line described in sec. 
3 and 4. 
He also wishes to thank all who were involved in 
organizing  the 
conferences 
{\em Constrained Dynamics and Quantum Gravity}, 
Sta. Marguerita, September 1996, Italy;   
{\em New Voices in Gravitation and Quantum Gravity}, Penn State, 
November 1996, USA and
{\em 18th Texas Symposium on Relativistic Astrophysics}, December 
1996, Chicago, USA, for providing a   delightful and 
enthusiastic atmosphere for discussions and research.  
In addition, he thanks B. Carr, M. MacCallum and 
J. Louko, Bei-Lok Hu for providing the opportunity 
to present this work as a  seminar at 
Queen Mary Westfield College and University of 
Maryland, respectively.

\end{document}